\documentclass[12pt]{article}
\usepackage[final]{graphics}
\usepackage{epsfig}
\usepackage{amsmath, amsthm, amssymb}
\usepackage{cite}
\usepackage{latexsym}
\usepackage{sparticles}
\usepackage{feynmf} 
\usepackage{hep}
 \newcommand {\nc} {\newcommand}
\nc {\tstrut} {\vline  height 2.5ex depth 1.7ex width 0ex}
\nc {\ustrut} {\vline  height 2.0ex depth 0.5ex width 0ex}
\numberwithin{equation}{section}
\begin{document}
\title{Polarisation observables in lepton antilepton to 
\\
proton antiproton reactions including lepton mass}
\author{N.~H.~Buttimore\,\,
and E.~Jennings\footnote{elise@maths.tcd.ie}
\\ \emph{School of Mathematics, Trinity College, Dublin, Ireland}}
%\date{ \ }
%
\maketitle
\abstract{
\noindent
General expressions, including the lepton mass, for the spin averaged differential cross section for the annihilation reaction lepton antilepton to proton antiproton are given, as well as general formulae for the single and double spin asymmetries in the centre of mass frame. In particular we discuss the single spin asymmetry, normal to the scattering plane, which measures the relative phase difference between nucleon electromagnetic form factors $G_E$ and $G_M$. Recent experimental investigations of these form factors in the space and time like region are reviewed. It is thought that measurements of the phase of these form factors will provide fundamental information on the internal nucleon structure. The phases between $G_E$ and $G_M$ are accessible through polarisation observables measured in the antiproton proton to lepton antilepton reaction, or in its time reversed process.}    
%
%\PACS{{12.20.Ds}{Specific calculations (QED)}\and {13.40.Gp}{Electromagnetic form factors} \and {annihilation}{Hadron production in e-e+ interactions} \and {13.75.Cs}{Nucleon-nucleon interactions
%(including antinucleons, deuterons, etc.)} \and {13.85.Fb}{ Inelastic scattering: two-particle final states} \and {14.20.Dh}{Protons and neutrons}}}
%
\section{Introduction}
\openup 1\jot
The form factors of hadrons as measured both in the space like and time like domains provide fundamental information on their structure and internal dynamics. As is well known, experiments on the scattering of high energy electrons by protons make it possible to determine the proton form factors in the region of space like momentum transfer ($q^2<0$). Information on the time like form factors is accessible through annihilation reactions such as $\bar{p}\,p \to l^+\,l^-$ where lepton, $l$, refers to the electron, muon or tau lepton.

At present there is great theoretical interest in the nucleon time like form factors due to recent experiments that have raised serious issues. Experimental values for the proton have been obtained over the last 50 years via electron proton scattering, often using the Rosenbluth separation technique \cite{Rosenbluth:1950}. The magnetic proton form factor, which is the dominant term in elastic ep scattering, has been measured at $q^2$ values up to 31 GeV$^2$ in the space like region \cite{Arnold:1986} and from $\bar{p}\,p$ or $e^-e^+$ annihilation up to $q^2 = 5.6$ \,GeV$^2$ \cite{Andreotti:2003}. 
They show that the magnetic form factor, $G_M$, follows approximately a dipole form $G_D = \left[1 + q^2/0.71\,(\mbox{GeV}^2)\right]^{-2}$, where $q^2$ is the four momentum transfer squared. 

Recent measurements \cite{Jones:2000,Punjabi:2005wq} of the electron to proton polarisation transfer in $e^-\,p \to e^-\,p$ scattering at Jefferson Laboratory show that the ratio of Sachs form factors $G_E(q^2)/G_M(q^2)$ is monotonically decreasing with increasing $q^2$ in strong contradiction with the $G_E/G_M$ scaling assumed in the Rosenbluth separation method.
Recall that in the expression of the differential cross section for the unpolarised elastic scattering of electrons on protons the coefficient $-q^2/4\,M^2$ in front of $G_M^2$ inhibits the contribution of $G_E^2$ to $d\sigma/d\Omega$ with increasing $q^2$. It is this fact which has led some to conjecture that data on $G_E$ in the space like region extracted by Rosenbluth technique may be unreliable and should be ignored in the global analysis \cite{Adamuscin:2005aq}. The Rosenbluth method may also be considered  incomplete in the space like region because of its sensitivity to uncertain radiative corrections, including two-photon exchange effects \cite{Gakh:2005hh}.
Unexpected results have been observed in the measurements of the proton form factors in the time like region also: the Fermilab E835 \cite{Andreotti:2003} measurements of $|G_M|$ of the proton at $q^2 = 11.63$ and $12.43\,$ GeV$^2$ have shown that $|G_M|$ in the time like region is twice as large as in the space like region.   
Although the space like form factors of a stable hadron are real, the time like form factors have a phase reflecting the final state interaction of the outgoing hadrons in a reaction such as $e^+\,e^- \to p\,\bar{p}$. Kaidalov et al. suggest that it is these final state interactions which are responsible for the enhancement of $|G_M|$ in the time like region \cite{Kaidalov:2000ca}.

It is as a result of these experiments that a precise separation of form factors and polarisation measurements is planned \cite{Barone:2005pu} at the future antiproton facility at GSI.
The Polarised Antiproton eXperiment (PAX) collaboration plans to produce polarised antiprotons by spin filtering with an internal polarised gas target. 
The origin of the unexpected $q^2$ dependence of the ratio $G_E/G_M$ of the proton, as observed at Jefferson laboratory, can be clarified by a measurement of their relative phase in the time like region, which discriminates strongly between the models for the form factor. As has been suggested in Ref.~\cite{Dombey:1969wk} this phase can be measured via single spin asymmetry in the annihilation $p\,\bar{p} \to e^-\,e^+$ on a transversely polarised target. 

The proposed measurement of this phase at PAX will also contribute to the understanding of the onset of the pQCD asymptotics in the time like region and will serve as a stringent test of dispersion theory approaches to the relationship between the space like and time like form factors. The double spin asymmetry will limit the relative phase ambiguity and allow independent $G_E$---$G_M$ separation, serving as a check on the Rosenbluth separation in the time like region.      
Despite the fundamental implications of the phase for the understanding of the connection between the space like and time like factors, such measurements have yet to be accurately obtained.  

In general the Dirac and Pauli form factors, $F_1$ and $F_2$ are analytic functions of $q^2$.
 They take real values in the space like region $q^2 < 0$ due to the hermiticity of the electromagnetic Hamiltonian. In the time like region the form factors are complex on the real $q^2$ axis above threshold due to the unitarity of the S matrix and the time reversal invariance of the theory.    
 The analytic structure and phases of the form factors in the time like regime are thus connected by dispersion relations to the space like regime\cite{Baldini:2001tt,Ioffe:1974,Hammer:2001rh}. 
These dispersion relations allow for a coherent description of the nucleon electromagnetic form factors over a large range of momentum transfer. Measurement of form factors is particularly important at high momentum transfer as it serves to test the predictions of perturbative QCD. 
The electric and magnetic form factors $G_E$ and $G_M$ are defined in the next section in terms of the Dirac and Pauli form factors, $F_1$ and $F_2$.

This paper is organised as follows: in section 2 we discuss the proton antiproton current in the time like region, given as a function of the electromagnetic form factors. We give the expression for the spin averaged differential cross section, including the lepton mass and show it reduces to the previously published result \cite{Dubnickova:1992ii} in which this mass has been neglected. In section 3 we present the centre of mass variables, derived by simple kinematics, which are used throughout this paper. In section 4 we introduce a scaled spin averaged differential cross section which will be used in the general expressions that follow. In section 5 and 6 we give expressions for the single and double spin polarisation observables, in the process $\bar{p}\,p \to l^-\,l^+$, including the lepton mass. It is shown that each polarisation expression reduces to previously published results when the lepton mass is neglected.   
\section{Unpolarised cross section} 

The differential cross section for proton-antiproton annihilation leading to a lepton-antilepton pair involves operators describing the proton and lepton currents, $J^{\mu}$ and $j_{\mu}$ respectively
\begin{eqnarray}
i\,e^2\,\bra{\bar{v}_{K'}} j_{\mu} \ket{u_{K}}
 \,\frac{1}{q^2}\,\bra{\bar{u}_{P}} J^{\mu} \ket{v_{P'}}
\end{eqnarray}
where $s = q^2 = q_{\mu}q^{\mu}$ is the square of the invariant 4-momentum in the s channel. Here we adopt the metric $q^2 = q^2_0 - \textbf{q}^2$ so that $q^2$ is positive in the time like region. This is given in terms of the final proton and antiproton momenta by
\begin{eqnarray}
q^{\nu} &=& P^{\nu} + P'\,^{\nu}\,.
\end{eqnarray}
If we assume the lepton has no structure we can replace $j_{\mu}$ by $-i\,e\,\gamma_{\mu}$. We shall look at the time reversed process $l^-\,l^+ \to p\, \bar{p}$. Using Lorentz and gauge invariance, the invariant amplitude due to one photon exchange is
\begin{eqnarray}
\cal{M} & = &
i\,e^2\,\bar{v}(K')\,\gamma_{\mu}\,u(K)\,\,\frac{1}{s}\,\,\bar{u}(P)\left[\,\gamma^{\mu}\,F_1(s) + \frac{i}{2\,M}\,\sigma^{\mu\nu}\,q_{\nu}\,F_2(s)\,\right]v(P')
\end{eqnarray}
where $M$ is the mass of the proton and $\sigma^{\mu\nu} = \frac{1}{2}\,i\,[\gamma^{\mu},\gamma^{\nu}]$. The momenta of the incoming lepton and antilepton are $K^{\mu}$ and $K'\,^{\mu}$ respectively and $P^{\nu}$ and $P'\,^{\nu}$ are the momenta of the outgoing proton and antiproton. These are given in the centre of mass frame in the next section.
\vspace{1ex}
\begin{figure}[!htb]
\begin{center}
\begin{tabular}{cccccccccccccccc}

\begin{fmffile}{one}    %one.mf will be created for this feynman diagram
\fmfframe(1,7)(1,7){    %Sets dimension of Diagram
\begin{fmfgraph*}(110,62)
%Sets size of Diagram
\fmfleft{i1,i2}
%Sets there to be 2 sources
\fmfright{o1,o2}
%Sets there to be 2  outputs
\fmflabel{$l^-(K)$}{i1}
%Labels one of the left sources
\fmflabel{$l^+(K')$}{i2}
%Labels one of the left sources
\fmflabel{$\bar{p}(P')$}{o1}
%Labels one of the right outputs
\fmflabel{$p(P)$}{o2}
%Labels one of the right outputs
\fmf{fermion}{i1,v1,i2} %Connects the sources with a vertex.
\fmf{fermion}{o1,v2,o2} %Connects the outputs with a vertex.
\fmf{photon,label=$\gamma(q)$}{v1,v2}
\fmfdot{v1}
\fmfblob{.16w}{v2}
%labels the conneting line.
\end{fmfgraph*}
 }
\end{fmffile}
\end{tabular}
\\
\caption{One photon exchange for $l^+\,+\,l^-\,\to\,p\,+\,\bar{p}$ in the $s$ channel.} \label{fey1}
\end{center}
\end{figure}
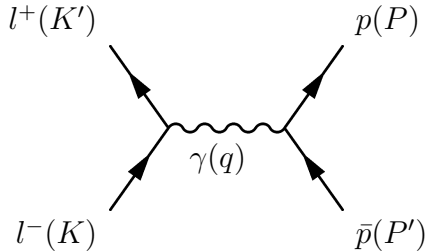

The Dirac and Pauli form factors, $F_1$ and $F_2$, are functions of $q^2$ and normalised at the origin such that $F_1(0) = 1$ and $F_2(0) = \mu_p - 1$ where $\mu_p$ is the magnetic moment of the proton. Using Gordon decomposition this becomes
\begin{eqnarray}
\cal{M} & = & i\,e^2\,\bar{v}(K')\,\gamma_{\mu}\,u(K)\,\frac{1}{s}\,\bar{u}(P)\,\left[\,\gamma^{\mu}\,G_M(s) + \frac{1}{2\,M}\,\left(\,P' - P\,\right)^{\mu}\,F_2(s)\right]\,v(P')\,.\nonumber
\end{eqnarray}
The Sachs electric and magnetic form factors are given by \cite{Sachs:1962}
\begin{eqnarray}
G_E(s) & = & F_1(s) + \frac{s}{4\,M^2}\,F_2(s)
\\ \nonumber 
\\
G_M(s) & = & F_1(s) + F_2(s)\,.
\end{eqnarray} 
By definition these form factors are equal at threshold ($s=4\,M^2$). In the Breit frame, $G_E$ and $G_M$ may be interpreted as the Fourier transforms of the charge and magnetisation distributions, respectively. We note here that while for t channel scattering ($q^2\,<\,0$) $G_E(q^2)$ and $G_M(q^2)$ are real on the negative real $q^2$ axis, the form factors can take complex values for $q^2 \,>\,4\,m_{\pi}^2$ corresponding to the annihilation reaction, where $m_{\pi}$ is the mass of the pion \cite{Hand:1963}.
 The unpolarised differential cross section for s channel annihilation of spin half particles in the centre of mass system is
\begin{eqnarray}
\frac{d\sigma}{d\Omega} & = & \frac{\beta}{64\pi^2 s}\,\frac{1}{4}\,\sum_{\mbox{\tiny spin}} |{\cal M}|^2 
\end{eqnarray}
\textwidth 14cm
where $\cal{M}$ is the invariant amplitude for the process and $\beta$ is a flux factor. In an annihilation reaction of two spinor particles of mass $m_i$ producing a pair of mass $m_f$ the flux factor $\beta$ is given by
\begin{eqnarray}
\beta& = &\left( \frac{s - 4\,m_f^2}{s - 4\,m_i^2}\right)^{1/2}.
\end{eqnarray}
In this case where all the particles are unpolarised it is necessary to sum over the polarisation of the final particles and average over the initial polarisations of the particles. 
The spin averaged differential cross section for $l^+ + l^- \to p + \overline{p}\,$  scattering in terms of Mandelstam varibles $s$ and $t$ is: 
%\begin{eqnarray}
%\label{unpol}
%\frac{d\sigma}{d\Omega}
%& =
%& \frac{\alpha^2}{2\,s^3}\frac{1}{\strut \sqrt{s - 4\,m^2}\, \sqrt{s - 4\,M^2}}\,
%\Bigg\{
%\,s^2\,\left(\,\tau - 1 \right)
%\,|G_{M}|^2 \,
%-\,2\,s\,m^2\,\left(\,|G_{M}|^2 - |G_{E}|^2\,\right)\nonumber \\[1ex]
%&&\qquad
% +\,2\,\left[\,\left(t\,-\,m^2 - M^2 \right)^2 + s\,t \right]\,\left(\,|G_{M}|^2\, \tau - |G_{E}|^2 \right)\,\Bigg\}
%
%\end{eqnarray}
%
%
\begin{eqnarray}
\label{unpol}
\frac{d\sigma}{d\Omega}
& =
&\alpha^2 \,\beta\,\frac{1}{s^3\,\left(s - 4\,M^2\right)}\,
\Bigg\{
\,\frac{s^2}{2}\,\left(\,s - 4\,M^2 \right)
\,|G_{M}|^2 \,
 -\,4\,s\,m^2\,M^2\,\left(\,|G_{M}|^2 - |G_{E}|^2\,\right)\nonumber \\[1ex]
&&\qquad
  +\,\left[\,\left(t\,-\,m^2 - M^2 \right)^2 + s\,t \right]\,\left(\,s\,|G_{M}|^2 - 4\,M^2|G_{E}|^2 \right)\,\Bigg\}
\end{eqnarray}
where $m$ is the mass of the lepton and $t = \left(P - K\right)^2$. The flux factor $\beta = \sqrt{s-4\,M^2}/\sqrt{s-4\,m^2}$. We can simplify this expression by neglecting the mass of the lepton. In both cases it transpires that if polarisations are not measured the terms corresponding to $G_E$ and $G_M$ contribute separately to the cross section.  
\begin{eqnarray}
\label{reduc}
\frac{d\sigma}{d\Omega} & = &\frac{\alpha^2}{4}\,\frac{\beta}{s}\,\Bigg\{(1+\cos^{2} \theta)\,|G_{M}|^{2} +  \frac{1}{\tau}\,\sin^{2}\theta\,|G_{E}|^{2}\Bigg\}
\end{eqnarray} 
where $\theta$ is the centre of mass scattering angle, given in Eq.~\ref{theta} for nonzero lepton mass. This formula was first obtained in Ref. \cite{Dubnickova:1992ii}. The absolute values of the form factors can be determined by a Rosenbluth separation technique. Measurements are made at at a number of angles $\theta$, and at a fixed value of the total energy $q^2$.

The total angular momentum for one photon exchange in $l^+\,+ \,l^- \to p \,+\,\bar{p}$  is $\emph{J} = 1$. If we consider C and P-invariances there are two allowed states for the $l^-\,l^+$ ($p\,\bar{p}$): $S = 1$, $L = 0$ and $S = 1$, $L = 2$ where $S$ is the total spin and $L$ is the orbital angular momentum. As a consequence the $\theta$-dependence of the spin averaged cross section is \cite{Tomasi-Gustafsson:2006mf}:
\begin{eqnarray}
\frac{d\sigma}{d\Omega} \,\simeq\, a(s) + b(s)\,\cos^2\theta
\end{eqnarray} 
where $a(s)$ and $b(s)$ are quadratic functions of $G_E(s)$ and $G_M(s)$. This result agrees with the spin averaged differential cross section in (\ref{reduc}). 
\section{Centre of mass variables}
The equations connecting the invariant Mandelstam variables with the practical ones of energy and angle can be derived by simple kinematics and are given below for $ \bar{p}\,p \to l^+\,l^-$ in the centre of mass system \cite{Barnes:1963}.
The four-momentum of the  proton and antiproton is specified by the centre of mass energy $E = \frac{1}{2}\sqrt{s}$ and a three-momentum, $\textbf{p}$, of magnitude $p$. The energy of each lepton is $E$ also. In the following equations $s$, $t$ and $u$ are Mandelstam variables with $u = 2\,m^2+2\,M^2 -s -t$.
\begin{eqnarray}
P^{\mu} &=& \left(E,\,0,\,0,\,p\right)
\\\nonumber
\\
P'\,^{\mu} &=& \left(E,\,0,\,0,\,-p\right)
\\\nonumber
\\
\cos\theta &=& \frac{t - u}{\tstrut \sqrt{s - 4\,m^2}\,\sqrt{s - 4\,M^2}}
\\\nonumber
\\
\label{theta}
\sin\theta &=& \frac{2\,\tstrut \left[4\,m^2\,M^2 - \left(t-m^2-M^2\right)^2 -s\,t\right]^{1/2}}{\tstrut \sqrt{s - 4\,m^2}\,\sqrt{s - 4\,M^2}}\,.
\end{eqnarray}
In the calculations that follow there are three polarisation orientations. These correspond to polarisation in three directions which are called longitudinal, in the scattering plane and normal but often denoted $z$, $x$ and $y$, respectively. In the scattering plane, ($x$), is perpendicular to the direction of the outgoing baryon. Longitudinal, ($z$), means parallel to the direction of the outgoing baryon. Normal, ($y$), refers to normal to the scattering plane in the direction of $\textbf{p}\times \textbf{k}$ where $\textbf{k}$ is the lepton momentum and $\textbf{p}$ is the proton momentum, with $x$, $y$ and $z$ forming a right handed coordinate system.  
The general formula for the spin polarisation four-vector $S^{\mu}$ is
\begin{eqnarray}
S^{\mu} &=& \frac{2}{M}\,\left(\,\textbf{p}\cdot\textbf{s},\;M\textbf{s} + \frac{\textbf{p}\cdot\textbf{s}}{E + M}\,\textbf{p}\right)
\end{eqnarray}
where $P^{\mu} = \left(E,\textbf{p}\,\right)$, $P^{\mu}P_{\mu} = M^2$, $S^{\mu}P_{\mu} = 0$ and $\textbf{s}\,^2 = 1/4$ for a spinor particle.
The three polarisation four-vectors in the $x$, $y$ and $z$ directions are  
\begin{eqnarray}
S^{\mu}_x &=& \,\left(\,0,\,1,\,0,\,0\,\right)\nonumber
\\[1ex]
S^{\mu}_y &=& \,\left(\,0,\,0,\,1,\,0\,\right)\nonumber
\\[1ex]
S^{\mu}_z &=& \,\left(\,p,\,0,\,0,\,E\,\right)/M\nonumber\,.
\end{eqnarray}
These spin four-vectors are normalised such that $S^{\mu}S_{\mu} = -1$. 
\section{Asymmetries with lepton mass}
We shall retain the lepton mass for the reaction $l^+\,l^- \to p\,\bar{p}$ and derive the general expressions for the single and double spin observables. It will be convenient to define a scaled unpolarised cross section given below in terms of $s$, $t$.
\begin{eqnarray}
\frac{d\sigma}{d\Omega}&=& \frac{\alpha^2}{4}\,\frac{\beta}{s}\,D
\end{eqnarray}
where D is given by:
\begin{eqnarray}
\label{D}
D
&=
&\frac{2}{(s/4\,M^2 - 1)s^2}\,
\bigg\{\,2\,
\left[\,\left(\,t\,- \,m^2 - M^2 \right)^2 + s\,t \right]\,\left(\,\frac{s}{4\,M^2}\,|G_{M}|^2 - |G_{E}|^2 \right)\nonumber \\[2ex]
&&\qquad
+ \,s^2\,\left(\,\frac{s}{4\,M^2} - 1\,\right)\,|G_{M}|^2 - 2\,s\,m^2\,\left(\,|G_{M}|^2 - |G_{E}|^2\right)\bigg\}\,.
\end{eqnarray}
We can also express $D$ in terms of the centre of mass scattering angle, $\theta$ and reduce it to the following:
\begin{eqnarray}
D
&=&
\left(\,1 - \frac{4\,m^2}{s}\right)\,\sin^2\theta\left[\,\frac{4\,M^2}{s}\,|G_E|^2 - |G_M|^2\,\right] + 2\left[\,\frac{4\,m^2\,M^2}{s^2}\,|G_E|^2 + |G_M|^2\right]\,.\nonumber
\\
\end{eqnarray}
%or
%\begin{eqnarray}
%D
%&=&
%\frac{4\,M^2}{s}\,\left[\,\frac{2\,m^2}{s} + \left(\,1 - \frac{4\,m^2}{s}\right)\,\sin^2\theta\right]\,|G_E|^2\,
%+ \left[\,2 - \left(\,1 - \frac{4\,m^2}{s}\right)\,\sin^2\theta\right]\,|G_M|^2\nonumber
%\end{eqnarray}
This general result for the spin averaged cross sections above reduces to Eq.~(\ref{reduc}) when the lepton mass $m \to 0$.

It has been noted in \cite{Tomasi-Gustafsson:2005kc} that it is possible to define an angular asymmetry, $\mbox{R}$, which can be measured from the differential cross section at $\theta = \pi/2$. This measurement does not require polarised particles and $\mbox{R}$ is defined in terms of the form factors as
\begin{eqnarray}
\mbox{R} &=& \frac{s\,|G_M|^2 - 4\,M^2\,|G_E|^2}{s\,|G_M|^2 + 4\,M^2\,|G_E|^2}\,.
\end{eqnarray}

In some cases it may be more convenient to use the invariant cross section $d\sigma/dt$ instead of the non-covariant cross section, $d\sigma/d\Omega$, given above. One cross section may be obtained from the other using the following equation in the centre of mass system.
\begin{eqnarray}
\frac{d\sigma}{d\Omega} &=& \frac{\sqrt{s - 4m^2}\sqrt{s - 4M^2}}{4\,\pi}\, \frac{d\sigma}{dt}\,.
\end{eqnarray}   
\section{Single spin asymmetry}
In order to determine the relative phase of the form factors it is necessary to perform experiments with polarised protons or antiprotons. When the antiproton in $l^+\,l^- \to p\,\bar{p}$ is polarised the differential cross section can be calculated using the following squared invariant amplitude:
\begin{eqnarray}
\sum_{\mbox{\tiny spin}}|{\cal M}|^{2} &&= \,\,\frac{e^4}{s^2}\, 
\mbox{Tr}\left(\cancel{K} + m\right)\gamma_{\mu}\left(\cancel{K'} - m\right)\gamma_{\nu} \\[2ex]
&&
\times\,\,\mbox{Tr}(\cancel{P} + M)\,\left[\gamma^{\nu}G_M + X^{\nu}F_2\right]
(\cancel{P'} - M)\left(1 + \gamma_5 \cancel{S}\right)
\left[\gamma^{\mu}G^*_M + X^{\mu}F^*_2\right]\,\nonumber
\end{eqnarray}
where $X^{\nu} = (P' - P)^{\nu}/2\,M$ and the  outgoing antiproton has polarisation vector $S^{\mu}$. We find in addition to the differential cross section Eq.~(\ref{unpol}) the following term. If the initial leptons are unpolarised we obtain one nonzero single spin asymmetry using the polarisation vector $S_{y} = (0,\,0,\,1,\,0)$.
\begin{eqnarray}
\label{AY}
\frac{\alpha^2}{16}\,\frac{\beta}{s^3}\ \,\frac{2\,\left( \epsilon_{\alpha\beta\gamma\delta}K'\,^{\alpha}P'\,^{\beta} + \epsilon_{\alpha\beta\gamma\delta}P'\,^{\alpha}K^{\beta}\right)P^{\gamma}S^{\delta} \,\left( u - t  \right)\,\mbox{Im}\,G_{E}\,G_{M}^*}{M\,\left( \tau - 1\right) }
\end{eqnarray} 
where $\tau = s/4\,M^2$. We can then evaluate the asymmetry parameter $A_y$, defined as a measure of the left-right asymmetry by
\begin{eqnarray}
A_y & = & \frac{\left(d\sigma/d\Omega\right)_{\uparrow}\, -\, \left(d\sigma/d\Omega\right)_{\downarrow}}{\left(d\sigma/d\Omega\right)_{\uparrow}\, +\, \left(d\sigma/d\Omega\right)_{\downarrow}}
\end{eqnarray}
where the subscripts $\uparrow$ and $\downarrow$ refer to the direction of the spin four-vector $S^{\mu}$ in the $y$ direction.   We then obtain the general ($m \not= 0$) single spin asymmetry for either a polarised proton or antiproton. 
\begin{eqnarray}
\label{y}
A_{y}
& = &
\left(\,1 - \frac{4\,m^2}{s}\,\right)\,\frac{2\,M\,\sin 2\theta\,}{\sqrt{s}\,D}\,\mbox{Im}\,G_{E}^*G_{M}
\end{eqnarray}
where $\theta$ is the centre of mass scattering angle, from Eq.~\ref{theta} a function of the lepton mass.
The predicted single spin asymmetry is substantial and has a distinct $q^2$ dependence which strongly discriminates between the analytic forms which fit the proton $G_E/G_M$
data in the space like region. As emphasised already by Dubnickova et al.~\cite{Dubnickova:1992ii}, the knowledge of the phase difference between the $G_E$ and $G_M$ may strongly constrain models for the form factors.

The existence of the T-odd single spin asymmetry normal to the scattering plane in the proton antiproton pair production requires a nonzero phase difference between the $G_E$ and $G_M$ form factors \cite{Rock:2001zi}. This observable is zero in the space like case. The complex phases of the form factors in the time like region make it possible for a single outgoing baryon to be polarised, even without polarisation in the initial state. By measuring the polarisation of one of the produced baryons it is possible to determine the phase of the ratio of form factors $G_E/G_M$ in the time like region.
As noted in \cite{Brodsky:2003gs} the factor $\sin 2\theta$ in $A_y$ causes this observable to vanish at $\theta = \pi/2$. Since the leptons in the initial state are unpolarised, at $\theta = \pi/2$ it is not possible to uniquely define the normal to the lepton-hadron plane; thus one cannot obtain a non-zero normal polarisation observable.

The other single spin observables $A_x$ and $A_z$ are nonzero only when the initial lepton is polarised. If the colliding leptons are longitudinally polarised there will be a contribution to the polarisation of the baryon produced. The observables $A_x$ and $A_z$ depend on the real phase of the form factors and the magnitude squared of $G_M$ respectively. Expressions for these two observables neglecting the lepton mass are available in Ref.~\cite{Dubnickova:1992ii}.
As we did in Section 2 for the unpolarised cross section, we can simplify Eq.~\ref{y} for $A_y$ by neglecting the lepton mass to obtain the following expression. This is in agreement with Ref.~\cite{Dubnickova:1992ii, Tomasi-Gustafsson:2005kc, Zichichi:1962}:
\begin{eqnarray}
A_{y} & = & \frac{\sin 2\theta\, \mbox{Im}\,G_{E}^*G_{M}}{\bigg\{(1+\cos^{2} \theta)\,|G_{M}|^{2} +  \frac{1}{\tau}\,\sin^{2}\theta\,|G_{E}|^{2}\bigg\}\sqrt{\tau}}\,\nonumber
\\ \nonumber
\\ \nonumber
\\
&=&\,\frac{(\tau - 1)\,\sin2\theta\,\mbox{Im}\,F_1F_2^*}{\bigg\{(1+\cos^{2} \theta)\,|G_{M}|^{2} +  \frac{1}{\tau}\,\sin^{2}\theta\,G_{E}|^{2}\bigg\}\sqrt{\tau}}\,.
\end{eqnarray} 
\section{Double spin asymmetries}
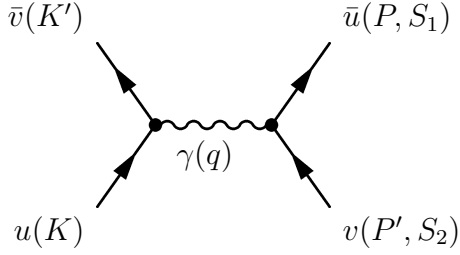
\begin{figure}[!htb]
 \begin{center}
 \begin{tabular}{cccccccccccccccc}
 \begin{fmffile}{oneb}    %one.mf will be created for this feynman diagram
 \fmfframe(1,7)(1,7){    %Sets dimension of Diagram
 \begin{fmfgraph*}(110,62)
  %Sets size of Diagram
  \fmfleft{i1,i2}
 %Sets there to be 2 sources
 \fmfright{o1,o2}
  %Sets there to be 2  outputs
 \fmflabel{$u(K)$}{i1}
  %Labels one of the left sources
 \fmflabel{$\bar{v}(K')$}{i2}
 %Labels one of the left sources
 \fmflabel{$v(P',S_2)$}{o1}
 %Labels one of the right outputs
 \fmflabel{$\bar{u}(P,S_1)$}{o2}
 %Labels one of the right outputs
 \fmf{fermion}{i1,v1,i2} %Connects the sources with a vertex.
 \fmf{fermion}{o1,v2,o2} %Connects the outputs with a vertex.
 \fmf{photon,label=$\gamma(q)$}{v1,v2}
 \fmfdot{v1}
 \fmfblob{.16w}{v2}
 %labels the conneting line.
 \end{fmfgraph*}
  }
 \end{fmffile}
 \end{tabular}
 \caption{Production of polarised protons and antiprotons in the $s$ channel.} \label{fey2}
 \end{center}
  \end{figure}
\noindent
If we now consider both the outgoing proton and antiproton as polarised with polarisation vectors $S^{\mu}_{1}$ and $S^{\mu}_{2}$ respectively we arrive at the following 
full expressions for the double spin observables $A_{xx}$, $A_{yy}$ and $A_{zz}$ including lepton mass
\begin{eqnarray}
\label{gen_xx}
A_{xx}
&=
&\emph{N}\,\Bigg\{\left[\,\left(\,t\,- \,m^2 - M^2 \right)^2 + s\,t \right]\,\left(\,s\,|G_{M}|^2 - 4\,M^2|G_{E}|^2 \right)\,
\\[1ex]\nonumber
&&
+\,\frac{1}{2}\,s\,(\,s - 4\,M^2\,)\, \left(\,s - 4\,m^2\right)\,\sin^2\theta\,|G_M|^2
- 4\,s\,m^2\,M^2\, \left(\,|G_{M}|^2 - |G_{E}|^2\right)\Bigg\}\nonumber
%
%\end{eqnarray}
%
\\[3ex]
%\begin{eqnarray}
\label{gen_yy}
A_{yy}
& =
& \emph{N}\,\Bigg\{\left[\,\left(\,t\,- \,m^2 - M^2 \right)^2 + s\,t \right]\,\left(\,s\,|G_{M}|^2 - 4\,M^2\,|G_{E}|^2 \right) \nonumber\\[1ex]
&&\qquad \qquad \qquad \qquad \qquad \qquad
\,- 4\,s\,m^2\,M^2\,\left(\,|G_{M}|^2 - |G_{E}|^2\right)
\Bigg\}
%
%
%\end{eqnarray}
\\[3ex]
%
%
%\begin{eqnarray}
\label{gen_zz}
A_{zz}
& =
& \emph{N}\,
\bigg\{\left[\,\left(\,t\,- \,m^2 - M^2 \right)^2 + s\,t \right]\,\left(\,s\,|G_{M}|^2 + 4\,M^2\,|G_{E}|^2 \right)\,\nonumber
 \\[1ex]
&&
+ \frac{1}{2}\,s^2\,(\,s - 4\,M^2\,)\, |G_{M}|^2 \, +\, 4\,s\,m^2\,M^2\,\left(\,|G_{M}|^2 - |G_{E}|^2\right)\bigg\}
\end{eqnarray}
where the coefficient $\emph{N}$ is given by
\begin{eqnarray}
\emph{N}& =& \frac{ 4}{s^2\,\left(\,s - 4\,M^2\,\right)\,D}
\end{eqnarray}
and $D$ is given in Eq.~(\ref{D}).
We can write this out explicitly as
\begin{eqnarray}
\emph{N}
& = &
\bigg\{\,
\left[\,\left(\,t\,- \,m^2 - M^2 \right)^2 + s\,t \right]\,\left(\,s\,|G_{M}|^2 - 4\,M^2\,|G_{E}|^2 \right) \\[2ex]
&&\qquad
 + \,\frac{s^2}{2}\,\left(\,s - 4\,M^2\,\right)\,|G_{M}|^2 - 4\,s\,m^2\,M^2\,\left(\,|G_{M}|^2 - |G_{E}|^2\right)\bigg\}^{-1}\,.\nonumber
\end{eqnarray}
 In these three equations there are many common factors with the unpolarised cross section as represented by $D$. These equations can be simplified to the following:
\begin{eqnarray}
D\,\left(\,1 - A_{xx}\,\right)
& = &
2\,|G_M|^2 - 2\, \left(\,1 - \frac{4\,m^2}{s}\,\right)\,\sin^2\theta\, |G_M|^2\,
\\[3ex]
%\end{eqnarray}
%
%
%\begin{eqnarray}
D\,\left(\,1 - A_{yy}\,\right)
& = &
2\, |G_M|^2
\\[3ex]
%\end{eqnarray}
%
%
%
%\begin{eqnarray}
D\,\left(\,1 - A_{zz}\,\right)
& = &
\frac{8}{(\,1-\,s/4\,M^2\,)\,s^2}\,
\bigg\{\left[\left(\,t\,- \,m^2 - M^2 \right)^2 + s\,t \right] \,|G_{E}|^2\,
\nonumber \\[1ex]
&&\qquad \qquad
+\,s\,m^2\left(\,|G_{M}|^2 - |G_{E}|^2\right)\bigg\}\,.
\end{eqnarray}
%
%All of the above formulae (\ref{gen_xx})-(\ref{gen_zz}) reduce to Eqs.~(\ref{xx})-(\ref{zz}) when we let the lepton mass $m \to 0$.\\
%
The last two double spin observables are $A_{xz}$ and $A_{zy}$. The expression for $A_{xz}$ is given in terms of invariant Mandelstam variables as:
\begin{eqnarray}
A_{xz} & = & \frac{\tstrut 2\,M\,\left[4\,m^2\,M^2 - (\,t - m^2 - M^2)^2 \,- s\,t\right]^{1/2}\,(t - u)}{\tstrut s^{5/2}\,\left(s-4\,M^2\right)\,D}\,\mbox{Re}\,G_{E}^*G_{M}\,.\nonumber
\\
\end{eqnarray}
We can simplify this expression greatly by using the  centre of mass scattering angle. We find $A_{xz}$ can be expressed as
\begin{eqnarray}
A_{xz} & = & \frac{2\,M}{\sqrt{s}\,D}\,\left(\,1 - \frac{4\,m^2}{s}\,\right)\,\sin 2\theta\,\mbox{Re}\,G_{E}^*G_{M}
\end{eqnarray}
and similarly the general expression for $A_{zy}$ is
\begin{eqnarray}
\label{gen_zy}
A_{zy} & = & \frac{2\,M}{\sqrt{s}\,D}\,\left(\,1 - \frac{4\,m^2}{s}\,\right)\,\sin 2\theta\,\mbox{Im}\,G_{E}^{*}G_{M}\,.
\end{eqnarray}
\\
The double spin asymmetry $A_{xy}$ is zero in the one photon exchange approximation although this is not true in general. For example when considering the two photon contribution mechanism this spin observable is non-zero \cite{Gakh:2005hh}. 

All of the above formulae (\ref{gen_xx})---(\ref{gen_zy}) reduce to previously published expressions \cite{Gakh:2005hh} for double spin observables when we let the lepton mass $m \to 0$.
In the equations that follow $\tau = s/4\,M^2$.
\begin{eqnarray}
A_{xy} & = & 0
\end{eqnarray}
\begin{eqnarray}
A_{xz} &=&\frac{\tau^{-1/2}\,\sin 2\theta\,\mbox{Re}\,G_{E}G_{M}^*}{(1+\cos^{2} \theta)\,|G_{M}|^{2} +  \frac{1}{\tau}\,\sin^{2}\theta\,|G_{E}|^{2}} 
\end{eqnarray}
\begin{eqnarray}
A_{zy} &=&\frac{-\tau^{-1/2}\,\sin 2\theta\,\mbox{Im}\,G_{E}G_{M}^*}{(1+\cos^{2} \theta)\,|G_{M}|^{2} +  \frac{1}{\tau}\,\sin^{2}\theta\,|G_{E}|^{2}}
\end{eqnarray}
\begin{eqnarray}
\label{xx}
A_{xx}
&=
&\frac{\sin^2\theta\,\left(\,\tau^{-1}\,|G_{E}|^2\, 
+ |G_{M}|^2
\,\right)}
{(1+\cos^{2} \theta)\,|G_{M}|^{2} +  \frac{1}{\tau}\,\sin^{2}\theta\,|G_{E}|^{2} 
}
\end{eqnarray}
\begin{eqnarray}
\label{yy}
A_{yy} &=& \frac{\sin^{2} \theta \,\left(\tau^{-1}\,|G_{E}|^{2} - |G_{M}|^{2} \right) \,
}{(1+\cos^{2} \theta)\,|G_{M}|^{2} +  \frac{1}{\tau}\,\sin^{2}\theta\,|G_{E}|^{2} 
}
\end{eqnarray}
\begin{eqnarray}
\label{zz}
A_{zz}
&=
&\frac{ \left(\,1 + \cos^2\theta\,\right)|G_{M}|^2\,  - \tau^{-1}\,\sin^2\theta\,|G_{E}|^2}
{(1+\cos^{2} \theta)\,|G_{M}|^{2} +  \frac{1}{\tau}\,\sin^{2}\theta\,|G_{E}|^{2} 
}\,.
\end{eqnarray}
As can be seen in the equations above, polarisation observables can be used to pin down the relative phases of the time like form factors. All of the double spin observables depend on the moduli squared of the form factors apart from $A_{xz}$ and $A_{zy}$ which contain the real and imaginary parts, respectively.  
\section{Conclusions}
The understanding of the electromagnetic structure of the nucleon, as revealed in proton antiproton reactions, is of upmost importance in any theory or model of strong interactions.
Abundant data over a large range of momentum transfer already exist and we have provided an overview of the r\^ole of the lepton mass in a study of the nucleon electromagnetic form factors in the time like region.

General expressions, including the lepton mass, for the spin averaged cross section as well as single and double spin asymmetries have been presented here.
For the annihilation reaction $\bar{p}\,p \to l^-l^+$ we simply invert the flux factor $\beta$ in Eq.~\ref{unpol} to obtain the unpolarised differential cross section:
\begin{eqnarray}
\frac{d\sigma}{d\Omega}
& =
&\frac{\alpha^2}{\beta}\,\frac{1}{s^3\,\left(s - 4\,M^2\right)}\,
\Bigg\{
\,\frac{s^2}{2}\,\left(\,s - 4\,M^2 \right)
\,|G_{M}|^2 \,
 -\,4\,s\,m^2\,M^2\,\left(\,|G_{M}|^2 - |G_{E}|^2\,\right)\nonumber \\[1ex]
 &&\qquad
 +\,\left[\,\left(t\,-\,m^2 - M^2 \right)^2 + s\,t \right]\,\left(\,s\,|G_{M}|^2 - 4\,M^2|G_{E}|^2 \right)\,\Bigg\}.
\end{eqnarray} 
Previously published results for polarisation observables in the positron electron to nucleon antinucleon reaction have been given in the case of zero lepton mass.
 Proposed polarised antiproton experiments by the PAX collaboration in the time like region will examine the moduli and relative phases of the form factors of the proton over a range of energies. Knowledge of these phases will make it possible to separate the magnetic and electric form factors in the time like region and thus permit significant tests of QCD and the asymptotic domain.
In the case of mu and tau final state pairs it will be necessary to retain the lepton mass in the formulae for both the spin averaged differential cross section and the polarisation observables in antiproton proton annihilation processes.
\section{Acknowledgements}
EJ would like to thank the Irish Research Council for Science, Engineering and Technology (IRCSET) for a postgraduate research studentship and Trinity College Dublin for the award of a Scholarship.

\end{document}